\documentclass[aps,prb,twocolumn,showpacs,preprintnumbers,amsmath,amssymb,superscriptaddress]{revtex4}
\usepackage{color}
\usepackage{graphicx}
\usepackage{dcolumn}
\usepackage{bm}

\usepackage[hypertex]{hyperref}
\newcommand{\nc}{\newcommand}
\nc{\be}{\begin{eqnarray}}
\nc{\ee}{\end{eqnarray}}
\nc{\bea}{\begin{eqnarray}}
\nc{\eea}{\end{eqnarray}}
\nc{\bean}{\begin{eqnarray*}}
\nc{\eean}{\end{eqnarray*}}
\nc{\mb}{\mbox}
\nc{\rnc}{\renewcommand}
\nc{\vk}{\mb{\bf k}}
\nc{\vp}{\mb{\boldmath$p$}}
\nc{\rr}{\mb{\boldmath$r$}}
\nc{\RR}{\mb{\boldmath$R$}}
\nc{\vz}{\hat {\mb{\bf z}}}
\nc{\vj}{\mb{\boldmath$j$}}
\nc{\vg}{\mb{\boldmath$g$}}
\nc{\vE}{\mb{\boldmath$E$}}
\nc{\vB}{\mb{\boldmath$B$}}
\nc{\vM}{\mb{\boldmath$M$}}
\nc{\vS}{\mb{\boldmath$S$}}
\nc{\x}{\mb{\boldmath$x$}}
\nc{\A}{\mb{\boldmath$A$}}
\nc{\va}{\mb{\boldmath$a$}}
\nc{\vq}{\mb{\boldmath$q$}}
\nc{\vn}{\mb{\boldmath$n$}}
\nc{\vs}{\mb{\boldmath$\sigma$}}
\nc{\vt}{\mb{\boldmath$\tau$}}
\nc{\vpi}{\mb{\boldmath$\pi$}}
\nc{\nab}{\bm{\nabla}}
\nc{\X}{\sf x}

\begin{document}

\title{
Electric Charging of Magnetic Textures 
on the Surface of a Topological Insulator
}

\author{Kentaro Nomura}
\affiliation{
Correlated Electron Research Group (CERG), RIKEN-ASI, Wako 351-0198, Japan
            }
\author{Naoto Nagaosa}
\affiliation{
Correlated Electron Research Group (CERG), RIKEN-ASI, Wako 351-0198, Japan
            }
\affiliation{
Cross-Correlated Material Research Group (CMRG), RIKEN-ASI, Wako 351-0198, Japan
}
\affiliation{Department of Applied Physics, The University of Tokyo, Hongo, Bunkyo-ku, Tokyo 113-8656, Japan}

\date{\today}

\begin{abstract}
A three-dimensional topological insulator manifests gapless surface modes, described by two-dimensional Dirac equation.
We study magnetic textures, such as domain walls and vortices, in a ferromagnetic thin film deposited on a three-dimensional topological insulator.
It is shown that these textures can be electrically charged, ascribed to the proximity effect with the Dirac surface states.
We derive a general relation between the electric and the magnetic charges.
As a physical consequence, we discuss domain wall motion driven by an applied electric field, which promises magnetic devices with high thermal efficiency.
\end{abstract}

\maketitle


{\it Introduction}-----
Topological insulators are new quantum states of condensed matter systems in which surface states resides in the bulk insulating gap.
\cite{review_TI1,review_TI2}
A three-dimensional (3D) $\mathbb{Z}_2$ topological insulator supports novel topologically protected spin polarized 2D Dirac fermions on its surface.\cite{review_TI1,review_TI2}
One of unique features is the universal quantized topological magnetoelectric (ME) effect.\cite{Qi_2008,Essin_2009}
Namely, the electromagnetic response of topological insulators is characterized by the axion electrodynamics, called the $\theta$ term\cite{Wilczek_1987};
$
 L_{\theta}=\int d^3\!x\ \Big(\frac{\alpha}{4\pi^2}\Big)\,\theta\, \vE\cdot\vB
$
with $\alpha=e^2/\hbar c$ and $\theta=\pi$, the possible nonzero value
in time-reversal invariant systems (mod $2\pi$).
Since the $\theta$ term is written as a total derivative when $\vE$ and $\vB$ are expressed in terms of vector potentials, salient effects can be seen when the system has a boundary.\cite{Qi_2008,Qi_2009}
 Indeed, this term describes the surface quantum Hall effect (QHE) that is related with an induced magnetization,
$
 \vM=\pm{\alpha}\vE.
$
Although induced magnetization (magnetic field) is too weak to be probed, it has lead to a new theoretical tread, namely novel interaction effects between topological surface states and magnetic systems.
\cite{Qi_2008,Qi_2008_1D,Qi_2009,Liu_2009,Guo_2010,Garate_2010,Yokoyama_2010,Garate_2010b,Nunez_2010,Burkov_2010}

In this work, we study theoretically the magnetic textures, such as domain walls and vortices, in a ferromagnetic thin film deposited on the surface of a topological insulator. It is found that magnetic textures interacting with the topological surface states are electrically charged.
A general relation between the electric and the magnetic charges is derived.
 As a consequence of the principal result, we propose a new 
mechanism of domain wall motion. Devices using this operating principle possibly have higher efficiency than conventional one based on current-driven domain wall motion\cite{Parkin_2008,review_spintronics}.

{\it Induced charge}-----
The surface electronic structure of a (pristine) topological insulator can be described, for the simplest case, by the massless Dirac-Rashba Hamiltonian\cite{review_TI1,review_TI2}
\bea
 {\cal H}_{\rm surface}=-i\hbar v_F\vs\!\times\!\vz\cdot\!(\nab-ie\A^{(e)})+eA_0^{(e)},
\eea
where $A^{(e)}_{\mu}$ is the vector potential for external electric and magnetic fields, $v_F^{}$
 is the Fermi velocity of the surface modes.
The surface is on the $xy$ plane ($z=0$). The bulk topological insulator is located on the $z<0$ side.

We consider a thin insulating ferromagnet consisting of localized spins $\vS=S\vn$ ($\vn$ a unit vector) 
on the top of the topological insulator ($z>0$).
In the functional integral approach, dynamics of local spins 
 in the ferromagnetic film is described by the Lagrangian
\bea
 L_{M} =
\int\! d^2x\ \rho_s \Big[
-i\hbar \dot{\vn}\!\cdot\!\langle \vn| \nab_{\vn}^{}|\vn\rangle
-{\cal H}_{M}\Big]
\eea
Here 
$
{\cal H}_{M} =
\frac{1}{2}JS^2
\sum_{a=1}^{3}|\nab n_a|^2
$
,
$|\vn\rangle$ is a spin coherent state\cite{Auerbach_textbook} of local spins,
$\rho_s=d/a^3$ is the sheet density of local spins,
 $a$ is the lattice constant, and $d$ is the width of the ferromagnetic thin film.
In an anisotropic ferromagnet, the energy of one local spin pointing in the easy and the hard axis are denoted by
$-\frac{1}{2}KS^2$ and $\frac{1}{2}K_{\!\perp}S^2$, respectively, to be taken into ${\cal H}_{M}$.

The proximity interaction between the local spins and  the surface Dirac fermions is described by
\bea
 {\cal H}_{\rm exc}=-\Delta\vn\cdot\vs
\eea
where $\Delta=\frac{1}{2}J'S\frac{d_{\rm exc}}{a^3}$, $J'$ is the exchange interaction between local spin and surface electron, and $d_{\rm exc}$ is the range of the exchange interaction.
A finite $z$ component $n_z$ opens up a gap in the Dirac surface dispersion, driving the surface states into the QH regime.\cite{Qi_2008}
In the following we set the Fermi level at the Dirac point  $E_F=0$, in the middle of the gap when it opens.
As we discuss later, when $E_F$ shifts slightly from the Dirac point, the results remain unchanged. From practical point of view, fine-tuning of $E_F$ is an important issue. Recent experimental efforts made it possible by the FET technique\cite{Jo} and by the magnetic doping (Mn and Fe)\cite{Chen_2010}. In the former case, an insulating layer contacting with a topological insulator would be replaced by a magnetic one. The latter situation itself could be a natural realization of proposed phenomena in this work.
The fermion sector including the exchange interaction, in the functional formalism, is given by
\bea
 L_F\!&=&\!\int d^2x\ \psi^{\dag}\Big(i\hbar\frac{\partial}{\partial t}
-{\cal H}_{\rm surface}-{\cal H}_{\rm exc}\Big)\psi \nonumber \\
\!&=&\!\int d^2x\ {\overline\psi}\Big[
i\hbar\sigma_z\Big(\frac{\partial}{\partial t}-ie{\cal A}_0\Big)
\nonumber \\ &&\qquad\qquad
-\hbar v_F\!\vs\!\cdot\Big(\nab+ie
{\mb{\boldmath${\cal A}$}}
\Big)
+\Delta n_z
\Big]\psi,
\label{L_F}
\eea
 where 
$
{\cal A}^{\mu}=(A_0^{(e)},A_x^{(e)}+a_x,A_y^{(e)}+a_y)
$, $\va=\frac{\Delta}{e v_F}\vz\!\times\!\vn$ and ${\overline{\psi}}\equiv\psi^{\dag}\sigma_z$.
Equation (\ref{L_F}) indicates that an electron on the surface feels an effective magnetic field $\nab\!\times\!\va=\frac{\Delta}{ev_F}\nab\!\cdot\vn$ and electric field $-\dot{\va}=-\frac{\Delta}{ev_F}\vz\!\times\!\dot{\vn}$.
As the electric charge density and the current are generated in the QH regime as $\rho_e=-\sigma_HB_z$ and $\vj_e=-\sigma_H\vz\times\vE$, above local spin gauge fields induce the electric charge and current as
\bea
\rho_e^{\rm ind}=-
\Big(\frac{\sigma_{H}\Delta}{e v_F}\Big)
\nab\!\cdot\vn
,\quad
\vj_e^{\rm ind}=
\Big(\frac{\sigma_{H}\Delta}{e v_F}
\Big)
\frac{\partial\vn}{\partial t},
\label{j_ind}
\eea 
where $\sigma_H$ is the quantized Hall conductivity.
Namely, spin textures with nonzero $\nab\!\cdot\vn$ interacting with surface electrons generate the electric charge on the surface.
These relations are the basic new findings in the present work.

The same results can be reached from the effective action in the functional formalism, given 
 by integrating out fermion fields as follows:\cite{Jackiw_1984,Garate_2010}
\bea
 Z&=&\int\!\!{\cal D}\vn{\cal D}{\overline \psi}{\cal D}\psi\,
\exp\!\Big[
\frac{i}{\hbar}\!\int \!dt
\Big(L_{M}[\vn]+L_{F}[\overline\psi,\psi,\vn]\Big)
\Big]\nonumber \\
&=&\int\!{\cal D}\vn\ \exp\!\Big[
\frac{i}{\hbar}\!\int \!dt
\Big(L_{M}[\vn]+L_{\rm CS}[{\cal A}_{\mu}]\Big)
\Big]
\eea
where
$
 L_{\rm CS}[{\cal A}_{\mu}]=-\int d^2x\,\frac{1}{2}\sigma_{H}\epsilon^{\mu\nu\rho}{\cal A}_{\mu}\partial_{\nu}{\cal A}_{\rho}
$
is the Chern-Simons term, which
 is devided into three terms as
$
L_{\rm CS}[{\cal A}_{\mu}]=L_{\rm CS}[A^{(e)}_{\mu}]+L_{\rm CS}[a_{\mu}]+L_{I}[A^{(e)}_{\mu},\vn]$.
The first term describes the electromagnetic response of the surface half-integer QHE with the Hall conductivity\cite{Jackiw_1984,Qi_2008}
$
\sigma_{H}=\frac{e^2}{2h}{\rm sgn}(n_z).
$
In the following, we assume that a weak magnetic field is applied in the $z$ direction so that $n_z>0$.
 The second term is the Chern-Simons term for the `gauge field' of local spins, which corresponds to the correction to the Berry phase term of local spins.\cite{Garate_2010,Yokoyama_2010}
The interaction between the local spins and the external electromagnetic field is described by
\bea
 L_I=-\int d^2\!x\ \sigma_{H}\,\epsilon^{\mu\nu\rho}\partial_{\nu}a_{\rho}A^{(e)}_{\mu}
\equiv-
\int d^2\!x\, j^{\mu}_{\rm ind}A^{(e)}_{\mu}.
\label{int-term}
\eea
The interaction term shows that  magnetic textures induce an electric charge and a current density $j^{\mu}_{\rm ind}=(\rho_e^{\rm ind},\vj_e^{\rm ind})$, given by Eq.(\ref{j_ind}).
We note that these relations automatically satisfy the continuity relation: $\partial\rho_e^{\rm ind}/\partial t+\nab\cdot\vj_e^{\rm ind}=0$. 

{\it Electromagnetic duality}-----
It is enlightening to define the magnetic charge induced by a texture by
$
 \rho^{\rm ind}_m\equiv -\nab\cdot\vM
$
where $\vM=-\gamma_m\hbar S\rho_s\vn$ is the magnetization density and $\gamma_m\equiv g\mu_B/\hbar$ is the gyromagnetic ratio. By comparing with Eq.(\ref{j_ind}), we obtain a general relation between the electric and the magnetic charges
\bea
  \rho^{\rm ind}_e= -\Big(\frac{\gamma_e}{\gamma_m}\Big)\ \rho^{\rm ind}_m,
\label{e-m}
\eea
where $\gamma_e\equiv \Delta\sigma_H/\hbar S\rho_s ev_F$.
Therefore a magnetic texture with nonzero $\nab\!\cdot\vn$ is a natural generalization of dyons
\cite{Witten_1979}, previously proposed as a surface effect of topological insulator.
\cite{Qi_2009}.
Namely, when a point-like electric charge is brought close to the surface of a topological insulator, a Hall current circularly flows around it, which in turn gives rise to a radial magnetic field. The source of the radial magnetic field can be regarded as an image magnetic monopole induced by the electric charge.\cite{Qi_2009}
By definition, a dyon is a particle or soliton with both magnetic and electric charge, corresponding to the bound state of the electric charge and its mirror image magnetic monopole, and also to the newly found magnetic textures with Eqs.(\ref{j_ind}) and (\ref{e-m}).
We note that image magnetic monopoles discussed in Ref.\onlinecite{Qi_2009} arise even in the conventional QH systems, where only a transverse Hall current is the ingredient.
On the other hand, the present results are characteristic to the Dirac surface states with the exchange coupling to localized spins.
In a typical situation\cite{note}, $J'/a^2\simeq 0.05[{\rm eV}]$ and $d/d_{\rm exc}\simeq 50$, applying an electric field for instance $E^{(e)}\simeq 10^6 [{\rm V/m}]$
 corresponds to a magnetic field $B^{(e)}=1[{\rm mT}]$.


The electromagnetic correspondence Eq.(\ref{j_ind}) implies a possibility of electric manipulation of the magnetic textures. In the rest of the paper, we analyze magnetic vortices and domain walls, as physical consequences of Eq.(\ref{j_ind}) and (\ref{e-m}).

\begin{figure}[b]
\begin{center}
\includegraphics[width=0.4\textwidth]{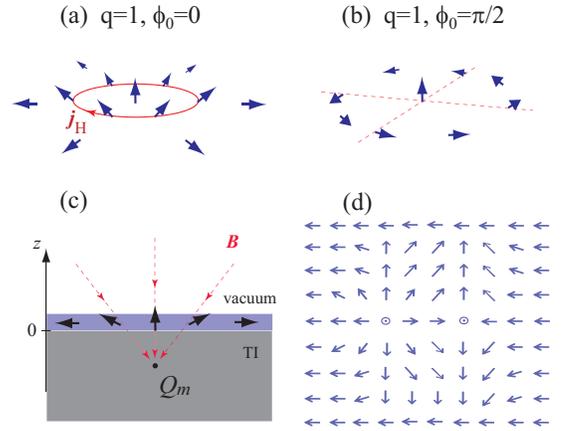}
\caption{
(a) Charged radial vortex and (b) neutral curling vortex are shown.
In the charged vortex, a circulating Hall current of the surface state is generated by the exchange interaction, which induces an electric charge.
(c) A magnetic monopole $Q_m$, which is a mirror image of a test electric charge, generates a radial vortex structure in the thin ferromegnet.
(d) Total charge of a vortex-antivortex pair vanishes.
}
\label{vortices}
\end{center}
\end{figure}

{\it Vortices}-----
We consider a vortex in a thin ferromagnet with a hard axis pointing in the $z$ direction, deposited on a topological insulator.
A solution for a vortex located at the origin can be written in the form;
$
 \vn=(\cos\phi\sqrt{1-n_z^2},\sin\phi\sqrt{1-n_z^2},n_z),
$
where $n_z$ depends only on $r=\sqrt{x^2+y^2}$, 
$
 \phi=q\tan^{-1}(y/x)+\phi_0
$,
$q$ is an integer, and $\phi_0$ is a constant.
Note that, for the vortex solutions,
$
 \nab\cdot\vn=\Big(\frac{1}{r}\frac{d}{dr}\Big[r\sqrt{1-n_z^2}\Big]
+\frac{q-1}{r}\sqrt{1-n_z^2}
\Big)\nonumber \\
\times
\cos\Big(\phi_0+(q-1)\tan^{-1}\frac{y}{x}\Big).
$
Vortices with $q=1$ are special in the sense that they have finite total charge:
$
 Q^{\rm ind}=\int d^2x \rho_e^{\rm ind}=-\frac{\Delta \sigma_{H}}{ev_F}2\pi R\cos\phi_0
$
in a disk geometry with a radius $R$.
For $q\ne 1$, on the other hand, the charge distribution is anisotropic around the vortex core, and the total charge vanishes.
Therefore, on the surface of topological insulators, vortex-antivortex symmetry (symmetry between $q=+1$ and $-1$) is electrically broken.

From microscopic point of view, this charging effect can be understood as follows.
The exchange interaction tends to align local spins in the ferromagnet and electron spins on the surface. The latter are related with the electric current as
$
\vj=-ev_F\vz\times\langle\psi^{\dag}\vs\psi\rangle.
$
For $q=1$ and $\phi_0=0$, as shown in Fig.1(a), the current flows circularly around the vortex core, and generates a Hall electric field
$\vE_{\rm ind}=\frac{1}{\sigma_{H}}\vz\times\vj$.
The electric charge of vortex with $q=1$ and $\phi_0=0$ stems from this induced electric field.
As $\phi_0$ increases from 0, the induced field diminishes and vanishes at $\phi_0=\pi/2$ (Fig.1.(b)).

Since a radial vortex on a topological insulator is electrically charged,
 it can be generated by an electric charge (and its mirror image magnetic monopole).
Specifically, a positive test electric charge, for instance, is screened by a vortex with negative electric charge, and a negative mirror magnetic monopole image is screened by induced positive magnetic charge as depicted in Fig.1(c).

When the system size is large enough, a single vortex cannot be generated below the Kosterlitz-Thouless temperature, $T_{KT}$; a vortex and an antivortex are bound.
When spins surrounding a vortex-antivortex pair point in the same direction as seen in Fig.1(d), the total charge vanishes.
A single vortex arises when $T>T_{KT}$ or when the system is small enough.

\begin{figure}[b]
\begin{center}
\includegraphics[width=0.3\textwidth]{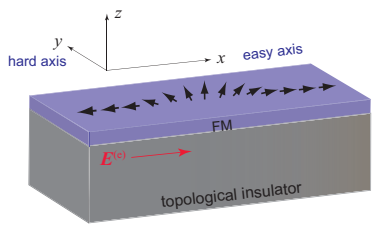}
\caption{
Neel domain wall structure in a ferromagnetic thin film, deposited on the top of a three dimensional topological insulator. The easy axis is in the $x$ direction, and the hard axis points in the $y$ direction. By an electric field, pointing in the $x$ direction drives the domain wall.
}
\label{dw}
\end{center}
\end{figure}

{\it Domain walls}-----
Finally we study so called a Neel domain wall, which corresponds to a projection of a radial vortex onto the $x$ axis as shown in Fig.2.
To describe domain walls, it is convenient using the notation:
$\vn=(\cos\theta,\sin\theta\sin\phi,\sin\theta\cos\phi)$.
We assume that an electric field and a magnetic field are in the $x$ direction,
the easy axis points in the $x$ direction, and the hard axis points in the $y$ direction, as illustrated in Fig.2.
The classical solution of a static domain wall of width  $\lambda=\sqrt{J/K}$ 
is given by
$
\phi=0,\ \ 
\theta=2\tan^{\!-1}\!\!\Big[\exp\!\big(-\frac{x\!-\!X}{\lambda}\big)\Big],
$
where $X$ is the center of the domain wall.
The induced electric charge is 
$
 Q_{DW}=\int d^2\!x\ \rho_{\rm ind}=L_y\frac{2\Delta\sigma_H}{ev_F^{}},
$
where $L_y$ is the width in the $y$ direction.
For $L_y=100[{\rm nm}]$, $Q_{DW}\sim 3e$.
As a consequence of this electric charging of domain walls, we expect that a domain wall behaves as a charged particle in an electric field.

To describe dynamics of domain walls, we study quantum fluctuations from above solution by regarding $X$ and $\phi$ as the dynamical valuables.
The Lagrangian for the domain wall is
\bea
 L_{DW}=\frac{\hbar SN_{DW}}{\lambda}\Big[\phi {\dot X}-\frac{\lambda SK_{\!\perp}}{2\hbar}\sin^2\!\phi
\nonumber \\
-\gamma_{e}E^{(e)}X
+\gamma_{m}B^{(e)}X
\Big],
\label{L_DW}
\eea
where $N_{DW}$ is the number of spins in the domain wall region.
The first term indicates that $\phi$ is momentum conjugate to $X$, while the second term describes the anisotropy effect.\cite{review_spintronics}
The third term is a new term derived from the effective interaction with the Dirac surface states Eq.(\ref{int-term}).

We note that under the correspondence: 
\bea
\gamma_e\leftrightarrow\gamma_m
\quad{\rm
and }\quad
E^{(e)}\leftrightarrow B^{(e)},
\eea
the third term in eq.(\ref{L_DW}) has a role of an external magnetic field, as described in the last term,
  that accelerates the domain wall in the conventional case\cite{Parkin_2008,review_spintronics}. 
This strong symmetry between electricity and magnetism is a marked character of the interface between a topological insulator and a thin ferromagnetic system.

As the domain wall velocity is given as 
$\frac{dX}{dt}=\frac{1}{\alpha_G}\lambda\gamma_mB^{(e)}$
under a magnetic field, below the Walker breaking field\cite{review_spintronics}, in the presence of an electric field,
\bea
 \frac{dX}{dt}=-\frac{1}{\alpha_G}\lambda\gamma_eE^{(e)}
\eea
for $E\le E_c\equiv 
K_{\perp}\frac{\alpha_G S}{2\hbar \gamma_e}
$.
Here $\alpha_G^{}$ is the Gilbert damping parameter.
With typical parameters
$v_F=3.0\times 10^5[{\rm m/s}]$, 
$a=0.5[{\rm nm}]$, $\lambda=100[{\rm nm}]$, 
$J'/a^2=0.05[{\rm eV}]$,
$K_{\perp}/(J'/a^2)=10^{-3}$, 
$d= 50[{\rm nm}]$ 
and
$\alpha_G=0.001$,\cite{review_spintronics,review_TI2},
the domain wall velocity is estimated as
$
 \big{|}\frac{dX}{dt} \big{|} \simeq   ({E}^{(e)}[{\rm V/m}]) \times 10^{-2}  [{\rm m/s}]
$
and
$E_c \sim
 10^6[{\rm V/m}]
$, where $({E}^{(e)}[{\rm V/m}])$ is the external electric field in units of ${\rm V/m}$.
On the other hand,  for $E> E_c$, the solution oscillates in time.

Here we compare this newly proposed mechanism of domain wall motion with the conventional one driven by spin-polarized current.\cite{review_spintronics}
For the conventional case, a ferromagnet consisting of localized spins and conduction electrons are considered.
A spin polarized current due to the conducting electron exerts a force on the domain wall via the exchange interaction.
When a current flows in a background magnetization $\vn$ varying in space, a slight directional mismatch between $\vn$ and polarized electron spin arises a spin transfer to the magnetization. For a spatially slowly-varying $\vn$, this spin transfer torque is expressed by ${\mb{\boldmath$T$}}_{e}=-({\mb{\boldmath$v$}}_s\!\cdot\!\nab)\vn$, where ${\mb{\boldmath$v$}}_s$ is the drift velocity of spin-polarized electrons.\cite{review_spintronics}

In the present case, there is a mass gap which is large ($\Delta\sim 0.05[{\rm eV}]$) in the vicinity of a domain wall, and thus a longitudinal transport current is strongly suppressed.
The mechanism of domain wall motion on a topological insulator is essentially different from conventional one. The origin of the torque in our mechanism is interplay of the exchange interaction and the topological surface states.
Namely, the vicinity of the domain wall is in the QH regime, and thus 
under an electric field $\vE^{(e)}$
the surface state has a finite magnetization given by $\langle\psi^{\dag}\vs\psi\rangle=\frac{1}{e}\vz\times \vj\propto\vE^{(e)}$ \cite{Qi_2008}
that tends to rotate and flip the local spin magnetization.\cite{Garate_2010,Yokoyama_2010} Consequently, an applied electric field puts pressure on the domain wall to move, as a magnetic field does.\cite{review_spintronics}.

The electric manipulation of domain wall motion has been aimed for application to spintronics devices where information is written electrically.
We note that the thermal efficiency of the domain wall motion studied in the present work is much higher than that in the conventional current-driven mechanism.
In the conventional case, Joule heat is released everywhere current flows.
In our mechanism, dissipation originates with Gilbert damping that arises only in the domain wall region. 
This will be a strong advantage for the application to magnetic devices.

We discussed a Neel domain wall. A Bloch wall, which has a finite mass in the domain region and  a mass kink corresponds to the domain wall, has been discussed on the 1D edge of a 2D topological insulator in Ref.\onlinecite{Qi_2008_1D}. Since the area of 1D edge region is tiny, motion of domain walls is hardly manipulated.
On the 2D surface of a 3D topological insulator, a Bloch wall does not have net electric charge, and thus does not move under an electric field.
A closely related work was reported in Ref.\onlinecite{Garate_2010},
in which a magnetization pointing in the $+\vz$ direction can be flipped to the $-\vz$ direction (or vice versa) by coupling with the topological surface. However to flip the magnetization, a large Hall current is needed that is order of the critical current above which the QHE is broken.\cite{Garate_2010}
Our operating principle of domain wall motion
can be used for spin flip, within $\pm\hat{\bf x}$ direction, which has much better efficiency than that proposed in Ref.\onlinecite{Garate_2010}.

In the above arguments we assumed that the Fermi level $E_F$ is zero.
If the Fermi energy is completely out of the QH gap ($|E_F|>\Delta$) the surface states are metalic and the induced electric charges can be screened, and thus above phenomena are strongly suppressed. On the other hand, when the Fermi level lies in the gap, then above results essentially remain unchanged.\cite{Jo} The effect of nonzero $E_F$ arises only away from the domain walls. However,
in the real materials there is potential modulation due to the randomness.
It has been known that all electrons are localized in the quantum Hall regime with broken time-reversal symmetry, and thus they are inert, no matter how the gap is small, while the quantized Hall currents are robust over disorder.\cite{Nomura_2008}

In summary, we clarified that magnetic textures interacting with topological surface states are electrically charged. A general relation between electricity and magnetism is found.
The proposed mechanism of domain wall motion in this work potentially have a great advantage for the application to nonvolatile memory devices.\cite{Parkin_2008}


This research is supported by MEXT Grand-in-Aid No.20740167, 19048008, 19048015 and, and 21244053, Strategic International Cooperative Program (Joint Research Type) from Japan Science and Technology Agency, and by the Japan Society for the Promotion of Science (JSPS) through its ``Funding Program for World-Leading Innovative R \& D on Science and Technology (FIRST Program)''.


\begin{thebibliography}{10}
\bibitem{note}
To enhance the effects, a small Fermi velocity is required. We assumed $v_F=4\times10^{5}{\rm [m/s]}$ for BiSe[\onlinecite{review_TI2}].
\end{thebibliography}
\end{document}